\documentclass{webofc}
\usepackage[varg]{txfonts}   
\begin{document}
\title{Accelerated cosmological expansion without tension in the Hubble parameter\footnote{Invited talk ICGAC-XIII and IK15, July 3-7, Seoul}}
 \subtitle{Fast evolution of the Hubble parameter $H(z)$}
 
\author{\firstname{Maurice }\lastname{H.P.M. van Putten}\inst{1}\fnsep\thanks{\email{mvp@sejong.ac.kr}} 
}

\institute{614 Yeongsil-Gwan, Department of Physics and Astronomy, Sejong University, 143-747 Seoul, South Korea}

\abstract{The $H_0$-tension problem poses a confrontation of dark energy driving late-time cosmological expansion 
measured by the Hubble parameter $H(z)$ over an extended range of redshifts $z$. Distinct values 
$H_0\simeq 73$ km\,s$^{-1}$Mpc$^{-1}$ and $H_0\simeq 68$ km\,s$^{-1}$Mpc$^{-1}$ obtain from surveys of the Local Universe 
and, respectively, $\Lambda$CBM analysis of the CMB. These are representative of accelerated expansion with
$H^\prime(0)\simeq0$ by $\Lambda=\omega_0^2$ and, respectively, $H^\prime(0)>0$ in $\Lambda$CDM,
where $\omega_0=\sqrt{1-q}H$ is a fundamental frequency of the cosmological horizon in a Friedmann-Robertson-Walker
universe with deceleration parameter $q(z)=-1+(1+z)H^{-1}H^\prime(z)$. Explicit solutions
$H(z)=H_0\sqrt{1+\omega_m(6z+12z^2+12z^3+6z^4+(6/5)z^5)}$ and, respectively, $H(z)=H_0\sqrt{1-\omega_m+\omega_m(1+z)^3}$ 
are here compared with recent data on $H(z)$ over $0\lesssim z \lesssim2$. The first is found to be free of tension
with $H_0$ from local surveys, while the latter is disfavored at $2.7\sigma$. A further confrontation obtains in galaxy dynamics by 
a finite sensitivity of inertia to background cosmology in weak gravity, putting an upper bound of $m\lesssim 10^{-30}$eV on the mass of dark matter.
A $C^0$ onset to weak gravity at the de Sitter scale 
of acceleration $a_{dS}=cH(z)$, where $c$ denotes the velocity of light, can be seen in galaxy rotation curves covering $0\lesssim z \lesssim 2$.
Weak gravity in galaxy dynamics hereby provides a proxy for cosmological evolution.
} 
\maketitle

\section{Introduction}\label{intro}
Estimates of the Hubble constant $H_0=H(0)$, where $H(z)$ denotes the Hubble parameter as a function of redshift $z$,
primarily derive from surveys of the Local Universe and fits of power spectra of the Cosmic Microwave Background 
(CMB) in the framework of $\Lambda$CDM. The results $H_0\simeq 73$ km\,s$^{-1}$Mpc$^{-1}$ and, respectively, 
$H_0\simeq 68$ km\,s$^{-1}$Mpc$^{-1}$ are distinct at a level of confidence better than $3\sigma$ \citep{free17}. 
This $H_0$-tension problem is interesting for its potential implications for dark energy density $\rho_\Lambda$,
beyond merely $\Lambda = 8\pi\rho_\Lambda>0$ inferred from a deceleration parameter
$q=-\ddot{a}a/\dot{a}^2=-1+(1+z)H^{-1}H^\prime$ satisfying
\begin{eqnarray}
q=\frac{1}{2}\Omega_M - \Omega_\Lambda < 0
\label{EQN_q1}
\end{eqnarray}
in $\Lambda$CDM based on a classical vacuum in three-flat Friedmann-Robsertson-Walker
(FRW) universe with line-element 
\begin{eqnarray}
ds^2=-dt^2 + a(t)^2(dx^2+dy^2+dz^2)
\label{EQN_FRW}
\end{eqnarray}
described by a scale factor $a(t)$, $H=\dot{a}/a$, evolved by general relativity with $\Lambda$ constant. Here, $\Omega_M$ and $\Omega_\Lambda$ refer to baryonic and dark matter density $\Omega_M$ and, respectively dark energy density normalization to closure density $\rho=3H^2/(8\pi G$), where $G$ denotes Newton's
constant.

While $q(0)<0$ appears relatively secure from surveys of the Local Universe, the relationship (\ref{EQN_q1}) derives from
classical general relativity, i.e., a covariant embedding of Newton's gravitational potential energy $U_N$ in geodesic motion in a metric of four-dimensional spacetime based on Einstein's principle of equivalence.\footnote{Equivalence of gravitational fields locally around non-inertial observers, whether arising from a massive object or arising from acceleration as seen by Rindler observers \citep[e.g][]{fey03}.} Applied to galaxy dynamics, we commonly preserve equivalence of geodesic motion to Newton's picture of force balance between gravitational and inertial forces with inertial mass $m$ equal to gravitating mass $m_0$, given by rest-mass energy $m_0c^2$, where $c$ denotes the velocity of light. In particular, the latter is assumed to be scale-free, i.e., $m=m_0$ is assumed to hold true at arbitrarily
small accelerations $\alpha$ conform Newton's second law (a proportional relation between force and acceleration). 
It has been suggested that perhaps the latter should be relaxed to account for anomalous galactic dynamics \citep{mil99,smo17}.

Our background cosmology introduces a de Sitter scale of acceleration $a_{dS}=cH$,
whose present value on the order of $1\AA\,$s$^{-2}$ is small but non-zero. If $a_{dS}$ breaks equivalence between $m$ and
$m_0$, galaxy dynamics is expected to be anomalous at distances in weak gravity, where
\begin{eqnarray} 
\alpha<a_{dS}.
\label{EQN_WG}
\end{eqnarray}
Astronomical evidence for general relativity at low accelerations is limited to verification of 
gravitational accelerations  $\alpha\gtrsim 10^{-6}$m\,s$^{-2}$ (with $m=m_0$), which leaves a window for anomalies in
galactic dynamics in weak gravity (\ref{EQN_WG}). Recently, we derived inertia in unitary holography \citep{van17a,van17b,van17c} with the property that $m<m_0$ at accelerations $\alpha < a_{dS}$ supported by high resolution data on galaxy rotation curves (Fig. 1).
\begin{figure}[h]
\centering
\includegraphics[scale=0.5]{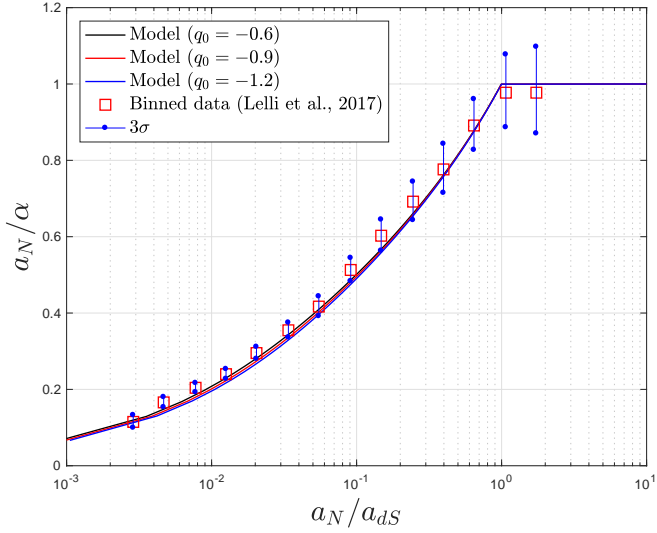}
\caption{High resolution data of \cite{lel17} on centripital accelerations $\alpha$ in galaxy rotation curves reveal an onset to weak gravity at $(a_N/a_{dS},\alpha/a_N)=(1,1)$ in transition to $m/m_0=a_N/\alpha<1$. This onset appears to be $C^0$, identified with a collusion of apparent Rindler and the cosmological horizon at Hubble radius $R_H$. Binned data shown are accompanied by $3\sigma$ uncertainties. Model curves (continuous lines) are included for various values of the deceleration parameter $q_0$, assuming a Hubble parameter $H_0 = 73$ km s$^{-1}$ Mpc$^{-1}$. (Reprinted from \cite{van17c}, data from \citep{lel17}).}
\label{fig-1}       
\end{figure}
With invariant kinetic energy $E_k$ and $U_N$ in orbital motion \citep{van17b}, this theory leaves the total energy $H=E_k+U_N$ and in particular the classical Lagrangian 
\begin{eqnarray}
L=E_k-U_N
\label{EQN_L}
\end{eqnarray}
unchanged. By volume, weak gravity makes up most of the Universe. {\em If inertia falls below its Newtonian value  
in weak gravity, then possibly the Hubble expansion is faster than what is expected in $\Lambda$CDM.} For this reason, 
anomalous galactic dynamics is a potential proxy of novel cosmological evolution.

In (\ref{EQN_FRW}), we have a cosmological horizon at the Hubble radius
\begin{eqnarray}
R_H = \frac{c}{H}.
\label{EQN_RH}
\end{eqnarray}
Defined as an apparent horizon in Cauchy surfaces of constant time $t$, these horizons are spheres with
area $A_H = 4\pi R_H^2$. As a compact surface, these horizons carry a finite fundamental frequency $\omega_0$
of an ordinary differential equation describing geodesic separation of associated null-geodesics,
\begin{eqnarray}
\omega_0 = \sqrt{1-q}H.
\label{EQN_om0}
\end{eqnarray}
 In cosmological holography, $\omega_0$ is picked up by the induced wave equation of massless fields, notably
 electromagnetic and gravitational fields, with dispersion relation 
 \begin{eqnarray}
 \omega = \sqrt{k^2 + \omega_0^2}
 \label{EQN_DR}
 \end{eqnarray}
for a frequency $\omega(k)$ with associated wave number $k$. Thus, the cosmological horizon induces
a dynamical dark energy 
\begin{eqnarray}
\Lambda=\omega_0^2
\label{EQN_LA}
\end{eqnarray}
which, in late time cosmology, is {\em inherently positive and small}. By (\ref{EQN_om0}), $\Lambda$ is 
dynamical and includes second time derivatives of $a(t)$. As such, including (\ref{EQN_LA}) in the 
FRW equation describing the Hamiltonian energy constraint,
\begin{eqnarray}
\Omega_M + \Omega_\Lambda = 1,
\label{EQN_HEC}
\end{eqnarray}
obtains an ordinary differential equations which is second order in time. It defines a 
{\em singular perturbation} of (\ref{EQN_HEC}) which, after all, is first order in time in 
$\Lambda$CMD. 

Here, we elaborate on accelerated cosmological expansion by (\ref{EQN_LA}) and in $\Lambda$CDM,
confronted with recent  Hubble data $H(z)$ \citep{sol17,far17} over an extended range of redshifts. This development is
facilitated by analytic solutions for both in late time cosmology, parameterized by $H_0=H(0)$
and $\omega_m=\Omega_M(0)$ of the Hubble parameter and density of (baryonic and dark) matter
at the present redshift $z=0$ (\S2). Our model for cosmological evolution has a Hubble parameter $H_0$ free 
of tension with estimates from surveys of the Local Universe A further confrontation with galaxy rotation curve data 
obtains in weak gravity at accelerations $\alpha<a_{dS}$ modeled by inertia of holographic origin (\S3). 
Our model identifies a holographic origin of dark energy and inertia, bringing together theory and data on 
cosmological evolution and anomalous galaxy dynamics (\S4). 

\section{Accelerated expansion in cosmological holography}\label{sec-1}

Evolution of the FRW scale factor $a(t)$ derives from (\ref{EQN_HEC}) with either (\ref{EQN_LA}) or $\Lambda$ constant
in $\Lambda$CDM. Parameterized by $H_0$ and $\omega_m$, the resulting Hubble parameter satisfies \citep{van17c}
\begin{eqnarray}
H(z)=H_0\sqrt{1+\omega_m(6z+12z^2+12z^3+6z^4+(6/5)z^5)}/(1+z),~~H^\prime(0) = H_0(3\omega_m-1)\simeq 0
\label{EQN_H1}
\end{eqnarray}
and, respectively,
\begin{eqnarray}
H(z)=H_0\sqrt{1-\omega_m+\omega_m(1+z)^3},~~H^\prime(0)=\frac{3}{2}\omega_mH_0\simeq 0.5H_0,
\label{EQN_H2}
\end{eqnarray}
where we used $\omega_m\simeq0.3$. According to (\ref{EQN_LA}), the Universe is presently at close to a minimum
value of $H(z)$, whereas $H(z)$ is decreasing to $H_0\sqrt{1-\omega_m}\simeq 0.83 H_0$ of a de Sitter Universe
in the distant future. This distinct behavior shows that, in late time cosmology, $H(z)$ will be larger for
(\ref{EQN_LA}) than in $\Lambda$CDM, the latter with a relatively stiff evolution by maintaining $H^\prime(z)>0$ well into the future.

Table 1 lists estimates of $(H_0,\omega_m)$ obtained by nonlinear model regression of (\ref{EQN_H1}-\ref{EQN_H2})
(Fig. 1) applied to recent data compilations of $(z_k,H(z_k))$. Fig. 1 includes distinct behavior (\ref{EQN_H1}-\ref{EQN_H2})) 
in the $qQ$-diagram, where $Q(z)=dq(z)/dz$. Table 1 includes estimates of $q_0=q(0)$ and $Q_0=Q(0)$ with $1\,\sigma$ 
uncertainties and fits to a cubic and quartic Taylor series expansion (with no priors on $q_0$ and $Q_0$) of $H(z)$.
\begin{figure}[h]
\centering
\includegraphics[scale=0.5]{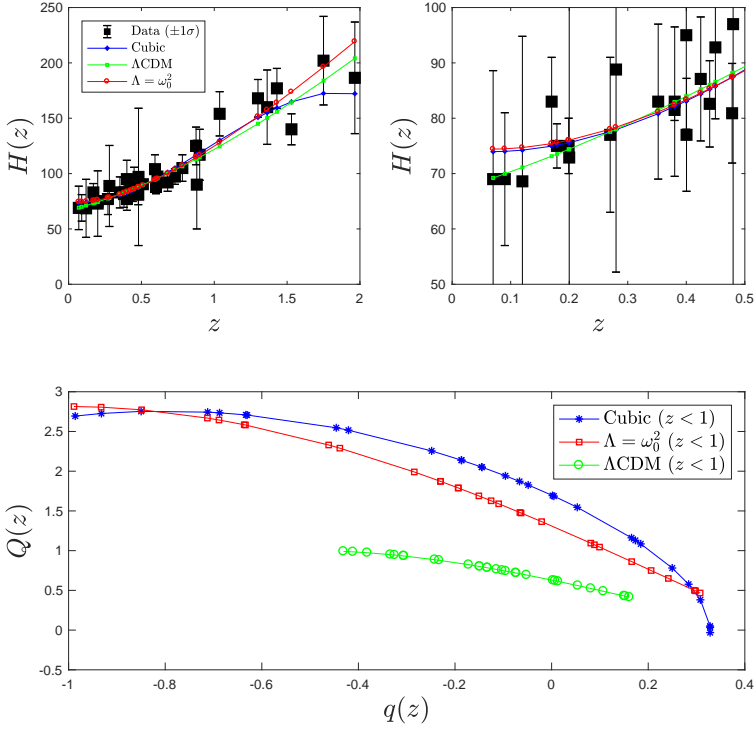}
\caption{(Top panels.) Fits of (\ref{EQN_H1}-\ref{EQN_H2}) and a Taylor series expansion to third order by nonlinear model regression to 
data $(z_k,H(z_k))$ over $0<z<2$. Note distinct behavior at $z\sim0$ with $H^\prime(0)\simeq0$ for (\ref{EQN_H1}) and $H^\prime(0)>0$
for (\ref{EQN_H2}). (Bottom panel.) In the $(q,Q)$-plane, results for (\ref{EQN_H1}) and consistent with the Taylor series expansion.
Results for $\Lambda$CDM are inconsistent with the latter at $2.7\,\sigma$. (Reprinted from \cite{van17c}, data from \citep{lel17}).}
\label{fig-2}       
\end{figure}

\begin{table}[h]
{\bf Table 1.} {Estimates of $(H_0,q_0,Q_0,\omega_m)$ with $1\,\sigma$ uncertainties by nonlinear model regression applied 
to the coefficients of the truncated Taylor series of cubic and quartic order, and to $(H_0,\omega_m)$ in
(\ref{EQN_H1}) with $\Lambda=\omega_0^2$ and (\ref{EQN_H2}) for  $\Lambda$CDM. $H_0$ is expressed in units of $\mbox{km~s}^{-1}\mbox{Mpc}^{-1}$.
(Reprinted from \cite{van17c}.)}
\center{\begin{tabular}{@{}lccccccc@{}}\mbox{}\\\hline\hline
	model & $H_0$ & $q_0$ & $Q_0$ & $\omega_m$ & $h^\prime(0)$ \\\hline
	Cubic                               & $74.4\pm4.9$  & $-1.17\pm0.34$  & $2.49\pm 0.55$ & - & -0.17 \\
	Quartic                             & $74.5\pm7.3$  & $-1.18\pm0.67$  & $2.54\pm 1.99$ &   & -0.18 \\
	$\Lambda=\omega_0^2$ & $74.9\pm2.6$  & $-1.18\pm0.084$ & $2.37\pm 0.073$ & $ 0.2719\pm0.028$ & -0.18 \\
	$\Lambda$CDM               & $66.8\pm1.9$  & $-0.50\pm0.060$ & $1.00\pm 0.030$ & $0.3330\pm0.040$ & 0.5 \\\hline
	\end{tabular}}\label{T1}\\
\end{table}
		
	Fig. 1 and Table 1 show a three-fold consistency among model-independent cubic and quartic fits and the model fit to
	(\ref{EQN_H1}). By cubic fit, $\Lambda$CDM is inconsistent with data at $2.7\,\sigma$. 
	
	 Here, $Q_0\simeq 2.5$ \citep{van15a} is representative a near-extremal value of $H(z)$ today. The associated relatively
	 high estimate of $H_0$ from the cosmological data $\{z_k,H(z_k)\}$ is free of tension with $H_0=73.24\pm1.74$km\,s$^{-1}$Mpc$^{-1}$ 
	 obtained from surveys of the Local Universe, providing quantitative support for a dynamic dark energy (\ref{EQN_LA}).
	 Combining results on $H_0$, we estimate \citep{van17c}
	\begin{eqnarray}
	H_0\simeq 73.75\pm 1.44\,\mbox{km\,s}^{-1}\mbox{Mpc}^{-1}.
	\label{EQN_H0a}
	\end{eqnarray}   
	
\section{Asymptotic behavior in weak gravity}

Applying (\ref{EQN_LA}) to (\ref{EQN_HEC}) yields a second order differential equation, we have
\begin{eqnarray}
q = \Omega_M - 2\Omega_\Lambda
\label{EQN_q2}
\end{eqnarray}
associated with $\Omega_\Lambda = (1/3)\left( 1-q \right)$ and $\Omega_M = (1/3) \left( 2 + q\right)$ and 
$w=(2q-1)/(1-q)$, defined by $p_\Lambda = w \rho_\Lambda$ between dark pressure and dark energy. In
the matter dominated era $q=1/2$ holds true in both (\ref{EQN_q1}) and (\ref{EQN_q2}), (\ref{EQN_LA}). In
late time cosmology $(q\lesssim -0.5)$, however, $q$ is {\em twice} the value (\ref{EQN_q1}) of  
$\Lambda$CDM, signifying {\em fast} evolution of the Hubble parameter $H(z)$.

The asymptotic regime of weak gravity $\alpha << a_{dS}$ satisfies the baryonic Tully-Fisher relation \citep{tul77} or,
equivalently, Milgrom's law \citep{mil83},
\begin{eqnarray}
\alpha = \sqrt{a_0 a_N}.
\label{EQN_AWG}
\end{eqnarray}
In a background cosmology with dark energy (\ref{EQN_LA}), we have \citep{van17c}
\begin{eqnarray}
a_0 = \frac{\omega_0}{2\pi}.
\label{EQN_a0}
\end{eqnarray}
By (\ref{EQN_om0}), (\ref{EQN_a0}) introduces a sensitivity of galaxy dynamics in the regime (\ref{EQN_AWG}) to
the cosmological parameters $(H,q)$, in addition to sensitivity to $a_{dS}$ at the onset to weak gravity ($a_N= a_{dS}$). 
A recent sample of galaxy rotation curves at intermediate redshifts $z\sim 2$ clusters close to the onset $a_N=a_{dS}$ but is 
in the weak gravity regime (\ref{EQN_WG}). The transition to (\ref{EQN_AWG}) is described by  
holographic inertia, sensitive to background parameters $(H,q)$. By (\ref{EQN_H1}), this theory accounts
for rotation curves from $z\sim0$ (Fig. 1) up to $z\sim2$ \citep{van17c}.

In this process, we encounter a deceleration parameter $q_0\simeq -1.18\simeq 1.6\times 10^{-10}$m\,s$^{-2}$ (Table 1)  with implied 
value $a_0\simeq 1.6\times 10^{-10}$m\,s$^{-2}$. In what follows, we consider the rate of decay to (\ref{EQN_AWG}) in
\begin{eqnarray}
\alpha = \sqrt{a_0a_N}\left(1+O\left( x^k\right)\right)~~(x<<1)
\label{EQN_gamma}
\end{eqnarray}
by specific values of $k>0$, where $x=a_N/a_0$.

Canonical estimates $a_0^\prime\simeq 1.2\simeq 1.6\times 10^{-10}$m\,s$^{-2}$ derive from fitting an 
interpolating function to rotation curve data such as Fig. 1 \citep[e.g.][]{mcg12}. Commonly used is
$f(x)=x/(1+x)$, $x=a/a_0^\prime$ satisfying $F_N=mf(x)\alpha$, gives 
\begin{eqnarray}
\alpha = \frac{1}{2}a_N\left( 1 + \sqrt{1+4x^{-1}}\right)=a_0^\prime\sqrt{x}\left( 1+\frac{1}{2}\sqrt{x}+\frac{1}{8}x+O\left(x^\frac{3}{2}\right) \right),
\label{EQN_a000}
\end{eqnarray}
showing $k=1/2$ in (\ref{EQN_gamma}).

Our value $a_0\simeq 1.6\times 10^{-10}$m\,s$^{-2}$ is defined by the asymptotic value of $\mu(y)=2\left<B(p)\right>_y$ in (\ref{EQN_a001}) in 
the notation of \citep{van17c}, 
\begin{eqnarray}
\alpha = \sqrt{\mu a_{dS}a_N},
\label{EQN_a001}
\end{eqnarray} 
defined as a thermal average of the ratio $B(p)$ of dispersion relations on the cosmological horizon and 3+1 spacetime within over 
momentum space as a function of $y=(a_N/a_{dS}) \left((1-q)/2\right)^{-\gamma}$,  $\gamma\simeq0.5$, whereby $y\simeq x/(\sqrt{2}\pi)$
in terms of $x$ above. With $u=p/T_H$, $T_H=(1-q)a_{dS}/(4\pi)$, $A=\Lambda/T_H = 16\pi^2/(1-q)$, we have
\begin{eqnarray}
\left<B(p)\right>_y = \frac{1}{W(y)} \int_0^\infty B(p) e^{-s^2} u^2du,~~W(y)= \int_0^\infty e^{-s^2} u^2du,~~s^2=\frac{u^2}{2\sigma^2},\label{EQN_a003}
\end{eqnarray}
where
\begin{eqnarray}
\sigma^2 \simeq \sqrt{A}\,y~~(u<<1).
\label{EQN_a004}
\end{eqnarray}
With $B(p) = 1+\frac{1}{2} u^2(1-1/A)+O(u^4)$, (\ref{EQN_AWG}) is reached with $\mu(y) = 1+ O\left(y\right)$, i.e., $k=1$ in 
(\ref{EQN_gamma}) of (\ref{EQN_a001}).

In a confrontation with data as shown in Fig. 1, we face the fact that $x$ remains finite. 
In comparing $a_0^\prime$ from a fit using an interpolation function with $a_0$ associated with (\ref{EQN_a0}), discrepancies
are expected by distinct decays $k=1/2$ and $k=1$ in (\ref{EQN_gamma}). A fit to a tail of data 
$y_0=a_N/a_{dS}=O\left(10^{-2}\right)$ of Fig. 1 involves the first two terms of (\ref{EQN_a000}), whereas (\ref{EQN_AWG}) is attained
to order $O(y)=O(x)$ in the latter. Writing $a_0^\prime = (1-\epsilon)a_0$, $a_0^\prime$ and $a_0$ satisfy
\begin{eqnarray}
\sqrt{a_0a_N}\left(1-\frac{1}{2}\epsilon\right) \left( 1+\frac{1}{2}\sqrt{x}+O\left(x\right)\right) = \sqrt{a_0a_N} \left( 1+ O\left(x\right)\right).
\end{eqnarray}
With $a_0=\omega_0/(2\pi)=\sqrt{1-q}H/(2\pi)=a_{dS}\sqrt{1-q}/(2\pi)$, we conclude that $a_0^\prime$ {\em under}-{estimates} $a_0$ by 
\begin{eqnarray}
\epsilon \simeq \sqrt{x} \simeq \sqrt{\frac{2\pi }{\sqrt{1-q}}}\sqrt{y_0} \simeq 21\% \sqrt{\frac{y_0}{10^{-2}}}.
\label{EQN_eps}
\end{eqnarray}
Indeed, a canonical value $a_0^\prime\simeq 1.18\times 10^{-10}$m\,s$^{-2}$ obtains by fitting the full expression (\ref{EQN_a000}) to the data of Fig. 1, whereas a fit of the asymptotic relation (\ref{EQN_AWG}) restricted to the tail $a_N/a_{dS}\lesssim 10^{-2}$ gives
\begin{eqnarray}
a_0\simeq 1.41\times 10^{-10}\mbox{m},\mbox{s}^{-2}
\label{EQN_a02}
\end{eqnarray}
larger than $a_0^\prime$ by about 19\%, as expected based on (\ref{EQN_eps}).

Our value $a_0\simeq 1.6\times 10^{-10}$m\,s$^{-2}$, obtained from (\ref{EQN_a0}) based on estimating the cosmological parameters $(H,q)$ in Table 1,
agrees with (\ref{EQN_a02}) to within about 17\%, which is within the $1\,\sigma$ (statistical and systematic) uncertainties of about 20\% in rotation curve data
\cite[e.g.][]{lel16a}.

\section{Conclusions and outlook}

The $H_0$ tension problem points to a discrepancy between accelerated expansion and relatively stiff
evolution in $\Lambda$CDM. We here present a dynamical dark energy based on a fundamental frequency of the cosmological
horizon, that is inherently positive and small. It introduces relatively fast evolution in the Hubble parameter today, satisfying
$H^\prime(0)\simeq0$ with $H_0$ larger than that expected in $\Lambda$CDM. A detailed confrontation with Hubble data
covering an extended range in redshifts obtains an estimate of $H_0$ (Table 1) in full agreement with $H_0$ obtained
from surveys of the Local Universe. Since (\ref{EQN_H1}-\ref{EQN_H2}) share the same parameters $(H_0,\omega_m)$ characterizing
late time cosmology,  dynamical dark energy and static dark energy can, for the first time,
be simultaneously compared with data. The results of Table 1 favor the first and disfavor the second by $2.7\,\sigma$.

Fast evolution (\ref{EQN_H1}) arises from novel behavior in the deceleration parameter $q(z)$, that changes the
Hamiltonian energy constraint (\ref{EQN_HEC}) to an ordinary differential equation which is second order in time,
rather than first order in time in $\Lambda$CDM. As such, (\ref{EQN_H1}) is a singular perturbation, disconnected
from $\Lambda$CDM. On this background, inertia of holographic origin is coevolving in the regime of weak gravity
(\ref{EQN_WG}) with a specific predictions for anomalous behavior in galaxy dynamics, whose asymptotic behavior
parameterized by (\ref{EQN_a0}) explicitly expresses sensitivity to background cosmology. 

Results on (\ref{EQN_a0}) derived from fitting (\ref{EQN_H1}) to cosmological data on the Hubble parameter and derived from a direct fit to
rotation curve data (Fig. 1) are consistent with rotation curve data within $1\,\sigma$ uncertainties. Our estimates of $a_0$ are
slightly higher than canonical estimates, that we identify with a relatively fast decay of (\ref{EQN_a001}) in our theory of weak
gravity to the asymptotic behavior (\ref{EQN_AWG}).

Conceivably, conditions of weak gravity might be reproduced in laboratory (or satellite) experiments. While we cannot
escape the presence of the gravitational field of the Earth (or the Sun), perhaps measurements on acceleration
along equipotential surfaces in the gravitational field of the Earth (or the Sun) can be realized to test for anomalies
$m<m_0$, by observing geodesic separation between particles in free fall, as an extension of Galileo's experiment.
Suitable accelerations below $a_{dS}$ may be imparted by gravitational or electrostatic forces.

{\bf Acknowledgements.} This research is supported in part by the National Research Foundation of Korea (No. 2015R1D1A1A01059793 and 2016R1A5A1013277).


\begin{thebibliography}{}
 \bibitem[Farooq et al.(2017)]{far17} Farooq, O., Madiyar, F.R., Crandall, S., \& Ratra, B., {\it ApJ}, {\bf 835}, 26 (2017) 
\bibitem[Feynman(2003)]{fey03} Feynman, R.P., Morimigo, F.B., \& Wagner, W.G., \textit{Feynman Lectures on Gravitation}  (Westview Press, Colorado, 2003), \S7.1-2
\bibitem{free17} Freedman, W.L.,  Nature Astronomy, \textbf{1}, 0121 (2017)
\bibitem[Lelli et al.(2016)]{lel16a} Lelli, F., McGaugh, S.S., \& Schombert, J.M., 2016, Astron. J., 152, 157
\bibitem[Lelli et al.(2017)]{lel17} Lelli, F., McGaugh, S.S., Schombert, J.M., \& Pawlowski, M.S., 2017, ApJ, 836, 152
\bibitem[McGaugh(2012)]{mcg12} McGaugh, S.S., 2012, ApJ, 143, 40
\bibitem[Milgrom(1983)]{mil83} Milgrom, M., 1983, ApJ 270, 365 
\bibitem[Milgrom(1999)]{mil99} Milgrom, 1999, M., Phys. Lett. A 253, 273
\bibitem[Smolin(2017)]{smo17} Smolin, L., 2017, arXiv:1704.00780
 \bibitem[Sol\'a et al.(2017)]{sol17} Sol\`a, J., G\'omez-Valent, A., \& de Cruz P\'erez, J., 2017, {\it ApJ}, {\bf 836}, 43 
 \bibitem[Tully \& Fisher(1977)]{tul77} Tully, R.B., \& Fisher, J.R., 1977, Astron. Astrophys. 54, 661 
\bibitem{van15a} van Putten, M.H.P.M., MNRAS \textbf{405}, L48 (2015)
\bibitem{van17a} van Putten, M.H.P.M., ApJ \textbf{837}, 22 (2017a)
\bibitem{van17b} van Putten, M.H.P.M., MPLA \textbf{32}, 1730019 (2017b)
\bibitem{van17c} van Putten, M.H.P.M., ApJ, to appear (2017c)
\end{thebibliography}
\end{document}